\documentclass[onecolumn,aps,preprintnumbers,amsmath,amssymb,superscriptaddress,nofootinbib]{revtex4}

\usepackage{graphicx,subfigure}
\usepackage{longtable}
\usepackage{dcolumn}
\usepackage{bm}
\usepackage[all]{xy}
\usepackage{color}
 \usepackage{graphicx}
\usepackage[usenames,dvipsnames]{xcolor}
\usepackage[unicode=true,pdfusetitle,
 bookmarks=true,bookmarksnumbered=false,bookmarksopen=false,citecolor=Turquoise,
 breaklinks=false,pdfborder={0 0 1},backref=false,colorlinks=true,pdfpagemode=FullScreen]
 {hyperref}

\usepackage{lscape}
\usepackage{tikz}
\usepackage{pgf}
\usepackage{color}
\usepackage{booktabs}
\usepackage{multirow}
\usepackage{siunitx}
\usepackage{setspace}
\usepackage{lipsum}

\makeatletter

\newcommand{\fmslash}[2][0mu]{%
  \mathchoice
    {\fmsl@sh\displaystyle{#1}{#2}}%
    {\fmsl@sh\textstyle{#1}{#2}}%
    {\fmsl@sh\scriptstyle{#1}{#2}}%
    {\fmsl@sh\scriptscriptstyle{#1}{#2}}}
\newcommand{\fmsl@sh}[3]{%
  \m@th\ooalign{$\hfil#1\mkern#2/\hfil$\crcr$#1#3$}}
\makeatother

\newcommand{\beq}{\begin{equation}}
\newcommand{\eeq}{\end{equation}}
\newcommand{\bea}{\begin{eqnarray}}
\newcommand{\eea}{\end{eqnarray}}
\mathchardef\minus="002D

\begin{document}
\title{Analytical Insights on Hadronic Top Quark Polarimetry}

\author{Zhongtian Dong}
\email{cdong@ku.edu}
\affiliation{Department of Physics and Astronomy, University of Kansas, Lawrence, KS 66045, USA}

\author{Dorival Gon\c{c}alves}
\email{dorival@okstate.edu}
\affiliation{Department of Physics, Oklahoma State University, Stillwater, OK, 74078, USA}

\author{Kyoungchul Kong}
\email{kckong@ku.edu}
\affiliation{Department of Physics and Astronomy, University of Kansas, Lawrence, KS 66045, USA}

\author{Andrew J.~Larkoski}
\email{larkoa@gmail.com}
\affiliation{Department of Physics and Astronomy, University of California, Los Angeles, CA 90095, USA \\
Mani L. Bhaumik Institute for Theoretical Physics, University of California, Los Angeles, CA 90095, USA}

\author{Alberto Navarro}
\email{alberto.navarro\_serratos@okstate.edu}
\affiliation{Department of Physics, Oklahoma State University, Stillwater, OK, 74078, USA}

\begin{abstract}
\noindent Top quark polarization provides an important tool for studying its production mechanisms, spin correlations, top quark properties, and new physics searches. Unlike lighter quarks, the top quark's polarization remains intact until its decay, enabling precise spin measurements. While the down-type fermions from $W$ boson decay are known to be effective spin analyzers, charged leptons have typically been the main target for most analyses. In this paper, we investigate the relevance of global jet dynamics -- considering kinematics, jet charges, and particle multiplicity -- for hadronic top quark polarimetry. The formalism used allows for analytical derivations obtained throughout the manuscript, offering deeper insights into the corresponding phenomenology.
\end{abstract}


\maketitle

\section{Introduction}

\noindent Top quark polarization is crucial for understanding its production mechanisms and shedding light into spin dynamics through observable decay patterns. Unlike lighter quarks, the top quark decays before its polarization is disrupted by soft QCD interactions, enabling precise spin measurements. Spin correlations between top quark pairs reveal complex structures, enhancing our understanding beyond event rates and making them useful in new physics searches~\cite{Bernreuther:1993hq,Arai:2007ts,Perelstein:2008zt,Krohn:2011tw,Barger:2011pu,Baumgart:2011wk,Buckley:2015vsa,Buckley:2015ctj,Goncalves:2016qhh,Goncalves:2018agy,Barger:2019ccj,Goncalves:2021dcu,Barman:2021yfh,MammenAbraham:2022yxp,Maltoni:2024tul}. 
Recent studies at the Large Hadron Collider (LHC) not only confirm these correlations~\cite{Mahlon:1995zn,Mahlon:2010gw,CDF:2010yag,D0:2011psp,ATLAS:2019zrq,CMS:2019nrx,Aguilar-Saavedra:2021ngj,Aguilar-Saavedra:2022kgy}, but also demonstrate that top quarks can be entangled in certain kinematic regimes~\cite{Afik:2020onf,Fabbrichesi:2021npl,Severi:2021cnj,Aguilar-Saavedra:2022uye,Afik:2022kwm,Afik:2022dgh,Severi:2022qjy,Aoude:2022imd,PhysRevD.109.115023,Han:2023fci,Cheng:2023qmz,ATLAS:2023fsd,Barr:2024djo,CMS-PAS-TOP-23-001,CMS:2024vqh,Dong:2024xsg,Cheng:2024btk}, motivating further research with increasing energy and luminosity to explore other quantum correlations that become more phenomenologicaly relevant at high energy scales, such as Bell's inequalities~\cite{PhysRevD.109.115023}.

The down-type fermion from the $W$ boson decay is the most effective spin analyzer for the top quark, providing particularly clean measurements in leptonic decays~\cite{Bernreuther:2010ny}. However, there is significant interest in measuring the polarization of top quarks that decay hadronically due to their sizable branching ratio. Conventionally, it is assumed that up to 50\% of the spin analyzing power can be restored in such scenarios using the softest of the two light jets from the top decay in the top quark rest frame~\cite{Jezabek:1994qs}. Ref.~\cite{Tweedie:2014yda} examined a straightforward method for analyzing spin in hadronic decays by using top quark decay kinematics, achieving a spin analyzing power of 64\% at leading order. This has been proposed as a promising kinematic method to probe entanglement and Bell's inequalities in the semileptonic channel of $t\bar t$ production at the LHC~\cite{PhysRevD.109.115023,Han:2023fci}. More recently, we have proposed a jet flavor tagging method and demonstrated the improvements to hadronic top quark polarimetry on simulated data using a Graph Neural Network~\cite{Dong:2024xsg}. We find that the spin analyzing power can be improved to approximately 0.75 (0.86), assuming an efficiency of 0.5 (0.2) for the network.

In this paper, we apply the general formalism to hadronic top quark polarimetry presented in Ref.~\cite{Dong:2024xsg} to investigate the relevance of global jet dynamics -- arising from kinematics, jet charges~\cite{Field:1977fa,Krohn:2012fg,Fraser:2018ieu,Kang:2023ptt}, and particle multiplicity -- on enhancing the sensitivity to hadronic top quark polarimetry. We aim to provide an analytical understanding for this subset of observables used in our previous machine learning study~\cite{Dong:2024xsg}.  
In Sec.~\ref{sec:framework}, we adopt an analytical approach, providing parametric estimates of the improvement with helicity angle, jet charge, and particle multiplicity. In Sec.~\ref{sec:results}, we study the phenomenological relevance of these observables, highlighting the phase space where they lead to substantial improvements. We conclude in Sec.~\ref{sec:conclusion}.
Appendix~\ref{app:impro} is reserved for a simple proof demonstrating that incorporating jet charge information always enhances the spin analyzing power. 

\section{Theoretical framework}
\label{sec:framework}
In this section, we explore the influence of helicity angle, jet charge, and particle multiplicity on the spin analyzing power in hadronic top quark decays, exploring the general formalism presented in Ref.~\cite{Dong:2024xsg}. We begin by reviewing the fundamental definitions for top quark polarization and jet charges. Following this, we apply the general formalism from Ref.~\cite{Dong:2024xsg} to the observables under consideration.

\subsection{Top Quark Polarization}

Due to its short lifetime, the polarization axis of the top quark $\vec{p}$, with  $0\leq |\vec{p\,}|\leq 1$, correlates with its decay products~\cite{Mahlon:2010gw,ParticleDataGroup:2022pth} as
\begin{align}
    \frac{1}{\Gamma}\frac{d\Gamma}{d\cos\theta_k}=\frac{1}{2}\left(1+\beta_k p\cos\theta_k \right)\,,
\end{align}
where $\cos\theta_k\equiv\hat {p}\cdot\hat{k}$ with $\hat k$ being the direction of a particular final state particle $k$ (calculated in the top rest frame), $p$ is the degree of polarization of the ensemble, $\beta_k$ is the spin analyzing power of the final state $k$, and $\Gamma$ is the top quark partial decay width. Among the top quark final states, the direction $\hat{d}$ of the down-type fermion from the $W$ decay displays maximal spin analyzing power due to the $V-A$ structure of the weak interactions, leading to $\beta_{\ell^+,\bar d}=1$. Whereas the charged lepton from the top quark decay can be uniquely identified, pinpointing the down-type quark from hadronic top quark decay in a collider environment is more challenging. Nevertheless, a proxy for the down-type quark direction can be defined by a probability weighted vector using a combination of kinematics and jet flavor tagging~\cite{Tweedie:2014yda,Dong:2024xsg}. Generally,  this direction can be defined as a combination of the soft $\hat q_\text{soft}$ and hard $\hat q_\text{hard}$ jet directions from the $W$ boson decay~\cite{Dong:2024xsg}:
\begin{align}
\label{eq:opt}    
\vec q_\text{opt} = \,\, p(\bar d\to q_\text{hard}|\{{\cal O}\}) \, \hat q_\text{hard} +p(\bar d\to q_\text{soft}|\{{\cal O}\}) \, \hat q_\text{soft}\,,
\end{align}
where $\{\cal O\}$ represents a set of observables, such as helicity angle, jet charge, and particle multiplicity. The spin analyzing power for this general direction is obtained from the vector's length $\beta_{\rm  opt}\equiv |\vec{q}_{\rm opt}|$, where $0\leq |\vec{q}_{\rm opt}|\leq 1$. In this article, we assume that the $b$-quark from the hadronic top decay has already been identified.   

\subsection{Jet Charge}

A definition of jet charge is given by~\cite{Field:1977fa,Kang:2023ptt}
\begin{align}
    {\cal Q}_{\kappa}\equiv \sum_{i\in J}\, z_{i}^{\kappa}Q_i\,,
    \label{eq:jetcharge}
\end{align}
where the sum runs over all particles $i$ in the jet $J$, $Q_i$ is the electric charge of particle $i$, $z_i$ is its the energy fraction of particle $i$ in a given jet, and $\kappa>0$ is a parameter introduced to ensure the infrared safety of the jet charge. Although the jet charge distribution can be computed from data, under a few assumptions, the functional form of the jet charge distribution conditioned on particle multiplicity can be approximated with a Gaussian function with appropriate mean and variance~\cite{Kang:2023ptt}. Therefore, we can estimate the impact of the jet charge in the spin analyzing power analytically. 
We note that the average charged particle multiplicity from on-shell $W$ decay has been measured to be $\langle N_\text{charged}\rangle = 19.39$, while the average charged pion multiplicity is $\langle N_{\pi^\pm}\rangle = 15.7$ \cite{ParticleDataGroup:2022pth}.

\subsection{Conditional Probabilities and Spin Analyzing Power}
To study the top quark polarimetry, we first determine the probability that the harder of the two jets from the $W$ decay, in the rest frame of the top quark, is initiated by a down-type quark. For concreteness,  we will consider the top quark decay $t\to b(W^+\to\bar{d}u)$ in the following discussion. This probability is calculated based on the measured values of a set of observables. In this paper, the observables we consider are:
\begin{itemize}
\item the cosine of the helicity angle $c_W$ (the angle between the down-type quark and opposite direction of the bottom quark in the frame of the $W$ boson), 
\item the jet charges of the hard and soft jets, ${\cal Q}_{\kappa,h}$ and ${\cal Q}_{\kappa,s}$, and 
\item particle multiplicity in the two jets, $N_h$ and $N_s$.
\end{itemize}

We start by considering only measurements of the helicity angle and jet charges, and work to simplify the relevant probability of Eq.~\eqref{eq:opt} following some reasonable assumptions.  For the case at hand, we want to compute the conditional probability
\begin{align}
p(\bar d\to q_\text{hard}|c_W,{\cal Q}_{\kappa,h},{\cal Q}_{\kappa,s}) = \frac{p(\bar d\to q_\text{hard},c_W,{\cal Q}_{\kappa,h},{\cal Q}_{\kappa,s})}{p(c_W,{\cal Q}_{\kappa,h},{\cal Q}_{\kappa,s})}\,.
\end{align}
The key assumption we make is that the jet charge measurements on the two $W$ decay product jets are independent of one another, and so the denominator factorizes to
\begin{align}
p(c_W,{\cal Q}_{\kappa,h},{\cal Q}_{\kappa,s}) &=p({\cal Q}_{\kappa,h},{\cal Q}_{\kappa,s}|c_W) \, p(c_W)=p({\cal Q}_{\kappa,h}|c_W) \, p({\cal Q}_{\kappa,s}|c_W) \, p(c_W)\,. 
\end{align}
We believe this is a reasonable assumption because only the sum of the mean values of the jet charges of the decay product jets is constrained to reproduce the electric charge of the $W$ boson.  The conditional probability can then be written as
\begin{align}
p(\bar d\to q_\text{hard}|c_W,{\cal Q}_{\kappa,h},{\cal Q}_{\kappa,s}) = \frac{p(\bar d\to q_\text{hard},{\cal Q}_{\kappa,h},{\cal Q}_{\kappa,s}|c_W)}{p({\cal Q}_{\kappa,h}|c_W) \, p({\cal Q}_{\kappa,s}|c_W)}\,.
\end{align}
 Going further, we decompose the numerator 
\begin{align}
\label{eq:prob1}
p(\bar d\to  q_\text{hard}|c_W,{\cal Q}_{\kappa,h},{\cal Q}_{\kappa,s}) = 
 \frac{p({\cal Q}_{\kappa,h},{\cal Q}_{\kappa,s}|c_W,\bar d\to q_\text{hard}) \, p(\bar d\to q_\text{hard}|c_W)}{p({\cal Q}_{\kappa,h}|c_W) \, p({\cal Q}_{\kappa,s}|c_W)}\,, 
\end{align}
where $p(\bar d\to q_\text{hard}|c_W)$ is the probability identified to define the optimal polarization axis using exclusively kinematic information based on Ref. \cite{Tweedie:2014yda}.
Since the probability in the numerator also factorizes assuming the independence of jet charge measurements, we have 
\begin{align}
p(\bar d\to q_\text{hard}|c_W,{\cal Q}_{\kappa,h},{\cal Q}_{\kappa,s}) = \frac{p({\cal Q}_{\kappa,h}|c_W,\bar d\to q_\text{hard}) \, p({\cal Q}_{\kappa,s}|c_W,\bar d\to q_\text{hard}) \, p(\bar d\to q_\text{hard}|c_W)}{p({\cal Q}_{\kappa,h}|c_W) \, p({\cal Q}_{\kappa,s}|c_W)}\,.
\end{align}
With a fixed flavor for the jet, the jet charge measurement is independent of kinematics, simplifying the numerator:
\begin{align}
 p(\bar d\to q_\text{hard}|c_W,{\cal Q}_{\kappa,h},{\cal Q}_{\kappa,s}) = 
 \frac{p({\cal Q}_{\kappa,h}|\bar d\to q_\text{hard}) \, p({\cal Q}_{\kappa,s}| u\to q_\text{soft}) \, p(\bar d\to q_\text{hard}|c_W)}{p({\cal Q}_{\kappa,h}|c_W) \, p({\cal Q}_{\kappa,s}|c_W)}\,. 
\end{align}

Next, we establish the probability $p({\cal Q}_{\kappa,h}|c_W)$, which can be defined by summing over all possible hard jets using the conditional probability
\begin{align}
p({\cal Q}_{\kappa,h}|c_W) = p({\cal Q}_{\kappa,h}|\bar d\to q_\text{hard}) \, p(\bar d\to q_\text{hard}|c_W) 
+p({\cal Q}_{\kappa,h}| u\to q_\text{hard}) \, p( u\to q_\text{hard}|c_W)\,, 
\end{align}
with a similar expression for $p({\cal Q}_{\kappa,s}|c_W)$. Thus, the conditional probability of interest is given by 
\begin{align}
p(\bar d\to q_\text{hard}|c_W,{\cal Q}_{\kappa,h},{\cal Q}_{\kappa,s}) &= \frac{p({\cal Q}_{\kappa,h}|\bar d\to q_\text{hard})}{p({\cal Q}_{\kappa,h}|\bar d\to q_\text{hard}) \, p(\bar d\to q_\text{hard}|c_W)+p({\cal Q}_{\kappa,h}| u\to q_\text{hard}) \, p(u\to q_\text{hard}|c_W)}\nonumber\\
&
\hspace{-0cm}\times\frac{p({\cal Q}_{\kappa,s}| u\to q_\text{soft})}{p({\cal Q}_{\kappa,s}|\bar d\to q_\text{soft}) \, p(\bar d\to q_\text{soft}|c_W)+p({\cal Q}_{\kappa,s}| u\to q_\text{soft}) \, p( u\to q_\text{soft}|c_W)}\\
&\hspace{-0cm}\times p(\bar d\to q_\text{hard}|c_W)\,. \nonumber
\end{align}

This result can be readily extended to include the measurements of particle multiplicity in the two jets, where
\begin{align}
&p(\bar d\to q_\text{hard}|c_W,{\cal Q}_{\kappa,h},N_h,{\cal Q}_{\kappa,s},N_s) 
= \frac{p({\cal Q}_{\kappa,h},N_h|\bar d\to q_\text{hard})}{p({\cal Q}_{\kappa,h},N_h|\bar d\to q_\text{hard})p(\bar d\to q_\text{hard}|c_W)+p({\cal Q}_{\kappa,h},N_h| u\to q_\text{hard})p( u\to q_\text{hard}|c_W)}\nonumber\\
&
\hspace{5.4cm}\times\frac{p({\cal Q}_{\kappa,s},N_s| u\to q_\text{soft})}{p({\cal Q}_{\kappa,s},N_s|\bar d\to q_\text{soft})p(\bar d\to q_\text{soft}|c_W)+p({\cal Q}_{\kappa,s},N_s| u\to q_\text{soft})p(u\to q_\text{soft}|c_W)}\nonumber\\
&\hspace{5.4cm}\times p(\bar d\to q_\text{hard}|c_W)\,.
\end{align}
If we assume that the quarks are first generation and isospin is an exact symmetry, then the particle multiplicity distribution of the $\bar d$ versus $u$ jets is identical, making the overall dependence on multiplicity distribution cancel out.  We then find
\begin{align}
p(\bar d\to q_\text{hard}|c_W,{\cal Q}_{\kappa,h},N_h,{\cal Q}_{\kappa,s},N_s) 
&= \frac{p({\cal Q}_{\kappa,h}|\bar d\to q_\text{hard},N_h)}{p({\cal Q}_{\kappa,h}|\bar d\to q_\text{hard},N_h)p(\bar d\to q_\text{hard}|c_W)+p({\cal Q}_{\kappa,h}| u\to q_\text{hard},N_h)p( u\to q_\text{hard}|c_W)}\nonumber\\
&
\times\frac{p({\cal Q}_{\kappa,s}| u\to q_\text{soft},N_s)}{p({\cal Q}_{\kappa,s}|\bar d\to q_\text{soft},N_s)p(\bar d\to q_\text{soft}|c_W)+p({\cal Q}_{\kappa,s}| u\to q_\text{soft},N_s)p( u\to q_\text{soft}|c_W)}\nonumber\\
&
\times p(\bar d\to q_\text{hard}|c_W)\,.
\label{eq14}
\end{align}
The universality assumption in the particle multiplicity between first and second generations is in general not accurate. However, given the limitations of collider experiments, it serves as a reasonable approximation for practical purposes. For instance, while we expect more kaons from the second generation quarks, achieving high accuracy in discriminating between kaons and pions is very challenging.  This universality assumption provides a conservative estimate on the conditional probability above, offering only minimal enhancement to top quark polarimetry. Any additional information would invariably lead to further improvements.\footnote{Discrimination between up-type and down-type jets can be enhanced by incorporating trajectory data from the tracking system, using charm tagging techniques~\cite{ATLAS:2018mgv,CMS-DP-2023-006}. We leave these improvements for the second-generation fermions from charm tagging for a future study.}

Finally, we observe that only the likelihood ratio of the jet charge distribution is necessary, where
\begin{align}
p(\bar d\to q_\text{hard}|c_W,{\cal Q}_{\kappa,h},N_h,{\cal Q}_{\kappa,s},N_s) 
&= \frac{1}{p(\bar d\to q_\text{hard}|c_W)+\frac{p({\cal Q}_{\kappa,h}| u\to q_\text{hard},N_h)}{p({\cal Q}_{\kappa,h}|\bar d\to q_\text{hard},N_h)}\,p( u\to q_\text{hard}|c_W)}\nonumber\\
& \label{eq:pdhard} 
\times\frac{1}{\frac{p({\cal Q}_{\kappa,s}|\bar d\to q_\text{soft},N_s)}{p({\cal Q}_{\kappa,s}| u\to q_\text{soft},N_s)}\,p(\bar d\to q_\text{soft}|c_W)+p(u\to q_\text{soft}|c_W)}\\
&
\times p(\bar d\to q_\text{hard}|c_W)\,. \nonumber
\end{align}
Notice that the jet charge dependence appears in two factors: one for the charge of the hard jet and another for the charge of the soft jet. The first factor depends on the likelihood ratio of the $u$ quark leading to the hard jet over the $\bar d$ quark resulting in the hard jet. For a fixed $N_{h}$, the mean of $p({\cal Q}_{\kappa,h}| u\to q_\text{hard},N_h)$ is slightly more positive than the mean of $p({\cal Q}_{\kappa,h}|\bar d\to q_\text{hard},N_h)$, see Eq.~\eqref{eq:means}. Hence, for large positive values of ${\cal Q}_{\kappa,h}$ the likelihood ratio $\frac{p({\cal Q}_{\kappa,h}| u\to q_\text{hard},N_h)}{p({\cal Q}_{\kappa,h}|\bar d\to q_\text{hard},N_h)}$ becomes very large and the right-hand side of Eq.~\eqref{eq:pdhard} approaches zero. This suggests that the average vector length should show an enhancement for large positive values of ${\cal Q}_{\kappa,h}$, resulting in a large spin analyzing power $\beta_{opt}$, see Eq.~\eqref{eq:opt}.  Using a similar argument, we can conclude that the ratio $\frac{p({\cal Q}_{\kappa,s}|\bar d\to q_\text{soft},N_s)}{p({\cal Q}_{\kappa,s}| u\to q_\text{soft},N_s)}$ becomes sizable for large negative values of ${\cal Q}_{\kappa,s}$, and we should expect an enhancement in spin analyzing power $\beta_{opt}$ for this limit as well. 

In the limit of a large multiplicity $N$ and assuming independent emissions, the general form of the jet charge distribution conditioned on multiplicity is Gaussian, which can be expressed as~\cite{Kang:2023ptt}
\begin{align}
\label{eq:gauss1}
p({\cal Q}_{\kappa}|\bar d,N) &= \frac{1}{\sqrt{2\pi \sigma^2}}\,e^{-\frac{({\cal Q}_\kappa-\mu_{\bar d})^2}{2\sigma^2}}\,, \\ 
\label{eq:gauss2}
p({\cal Q}_{\kappa}| u,N) &= \frac{1}{\sqrt{2\pi \sigma^2}}\,e^{-\frac{({\cal Q}_\kappa-\mu_{u})^2}{2\sigma^2}}\,.
\end{align}
Assuming that all particles in the jets are only pions and exact isospin conservation holds
(see comments on this assumption below Eq.~\eqref{eq14}), we have 
\begin{align}
\label{eq:means}
	\mu_{\bar d} &= \frac 1 3 N^{-\kappa}\left(1+\frac \kappa 2 (\kappa-1)\sigma_{z}^{2}N^{2}+\cdots \right), \nonumber\\  
 \mu_{u} &= \frac 2 3 N^{-\kappa}\left(1+\frac \kappa 2 (\kappa-1)\sigma_{z}^{2}N^{2}+\cdots\right),  \\
	\sigma^{2} &= \frac 2 3 N^{1-2\kappa}\left(1+\kappa(2\kappa-1)\sigma_{z}^{2}N^{2}+\cdots\right)\,.\nonumber
\end{align}
Since the soft and hard jets can display different particle multiplicities, $N_{s}$ and $N_{h}$, the means and variance would differ depending on whether the up/down quarks are identified as the soft/hard jets. In general, this implies that $\sigma_{z}^{2}N^{2}$ could vary for each of these cases. For simplicity, we assume that this term is the same for all distributions.

These expressions explicitly depend on the energy weighting exponent $\kappa$ from the definition of the jet charge, see Eq.~\eqref{eq:jetcharge}.  We have included contributions through the variance $\sigma_{z}^{2}$ of the multiplicity-conditioned energy fraction distribution $p(z|N)$, where we note that
\begin{align}
\langle z\rangle &= \int dz\, z\, p(z|N) = \frac{1}{N}\,, \\
\sigma_z^2 &= \int dz\, \left(
z-\frac{1}{N}
\right)^2\, p(z|N)\,.
\end{align}
Although $\sigma_{z}^{2}$ is in principle unknown, it is strictly non-negative, bounded, and can be assumed to scale as $\sigma_{z}^{2}\sim 1/N^{2}$~\cite{Kang:2023ptt,Kang:2023zdx}. 

\section{Results}
\label{sec:results}

While the evaluation of the mean spin resolving power must be performed numerically, the mean of the square resolving power can be expressed in closed form.  When using exclusively kinematic information, the square spin resolving power is given by 
\begin{align}
|{\vec q}_\text{opt}^\text{\hspace{-0.01cm} kin}|^2=1-2\,p(\bar d\to q_\text{hard}|c_W)p(\bar d\to q_\text{soft}|c_W)\left(
1-\hat q_\text{hard}\cdot  \hat{q}_\text{soft}
\right)\, . 
\end{align}
Here, the angle between the hard and soft jets in the top rest frame can be expressed in terms of the helicity angle $\theta_W$ as 
\begin{align}
1-\hat q_\text{hard}\cdot \hat q_\text{soft} = \frac{2}{1+\frac{(m_t^2-m_W^2)^2}{4m_t^2m_W^2}\,\sin^2\theta_W}\,,
\end{align}
where $m_t$ and $m_W$ are the masses of the top quark and $W$ boson, respectively. In this estimation, we neglect the effects of the $b$-quark mass.  The mean of the square resolving power is then obtained by integrating $ |\vec q_\text{opt}^\text{\hspace{0.007cm} kin}|^2$ over the distribution of the helicity angle, $p(c_W)$, 
\begin{align}
    \label{eq:qoptkin}
    \langle|\vec q_\text{opt}^\text{\hspace{0.007cm} kin}|^2\rangle &= \int\,dc_W\,p(c_W)\,|\vec q_\text{opt}^\text{\hspace{0.007cm} kin}|^2\,,
\end{align}
where the helicity angle distribution $p(c_{W})$ is defined as~\cite{Bern:2011ie}
\begin{align}
    \label{eq:polar}
    p(c_{W}) = \frac 3 8 f_{R}(1 \pm c_{W})^{2} + \frac 3 4 f_{0}(1-c_{W}^{2}) + \frac 3 8 f_{L}(1 \mp c_{W})^{2} \, . 
\end{align}
The upper sign corresponds to $W^+$ and the lower sign to $W^-$. The polarization fractions can be obtained from the following relations
\begin{align}
f_0 =2-5\langle c_{W}^{2}\rangle,\hspace{1cm}
f_L=-\frac{1}{2}\mp \langle c_{W}\rangle+\frac{5}{2}\langle c_{W}^{2}\rangle, \hspace{1cm}
f_R=-\frac{1}{2}\pm \langle c_{W}\rangle+\frac{5}{2}\langle c_{W}^{2}\rangle \, ,
\label{eq:polfrac}
\end{align}
where the fractions for the zero helicity, left-handed helicity, and right-handed helicity for the $W^+$ boson in the top quark rest frame are given by $f_0\approx 0.7$, $f_L\approx 0.3$, and $f_R=0$, respectively. For the $W^-$ decay, the quoted values for $f_R$ and $f_L$ are exchanged, as can be inferred from Eq.~\eqref{eq:polfrac}.

The integral from Eq.~\eqref{eq:qoptkin} evaluates to
\begin{align}
\langle |\vec q_\text{opt}^\text{\hspace{0.007cm} kin}|^2\rangle  &= \frac{m_t^4-5m_t^2m_W^2-2m_W^4}{m_t^4+m_t^2m_W^2-2m_W^4} 
+ \frac{12m_t^2m_W^4}{(m_t^4+m_t^2m_W^2-2m_W^4)\sqrt{m_t^4-m_W^4}}  \left(\tanh^{-1}\frac{\sqrt{m_t^2-m_W^2}}{\sqrt{m_t^2+m_W^2}}\right)\,. 
\end{align}

Plugging in the values for $m_{t}=172.57\,$GeV and $m_{W}=80.3692\,$GeV~\cite{ParticleDataGroup:2022pth}, we obtain 
\begin{align}
\sqrt{\langle |\vec q_\text{opt}^\text{\hspace{0.007cm} kin}|^2\rangle}\approx 0.640\,,
\end{align}
which is slightly larger than the mean spin resolving power $\langle |\vec q_\text{opt}^\text{\hspace{0.007cm} kin}|\rangle\approx 0.638$. Hence, $\sqrt{\langle |\vec q_\text{opt}^\text{\hspace{0.007cm} kin}|^2\rangle}$ also works as a good proxy for $\beta_{\rm opt}$, and we shall use it to numerically estimate the effect of the jet charge on the spin analyzing power. This is done by generalizing Eq.~\eqref{eq:qoptkin} to account for the average of $|\vec{q}_{\text{opt}}|^{2}$ over the charge of the hard and soft jets, and over the (cosine of the) helicity angle for a fixed particle multiplicity\footnote{In Appendix~\ref{app:impro}, we prove, using these results and general arguments, that the spin analyzing power is enhanced by incorporating jet charge information.} 
\begin{align}
    \label{eq:qoptmeansq}
    \langle |\vec{q}_{\text{opt}}|^{2} \rangle = \int \, dc_{W}\, p(c_{W}) 
     \int \, d{\cal Q}_{\kappa,h}\, d{\cal Q}_{\kappa,s} \,p({\cal Q}_{\kappa,h},{\cal Q}_{\kappa,s}|c_W,N_h,N_s)\,|\vec{q}_{\text{opt}}|^{2}, 
\end{align}

Taking into account the particle multiplicity, the joint probability distribution $p({\cal Q}_{\kappa,h},{\cal Q}_{\kappa,s}|c_{W},N_h,N_s)$ becomes 
\begin{align} 
p({\cal Q}_{\kappa,h},{\cal Q}_{\kappa,s}|c_W,N_{h},N_{s})&=\left[
p({\cal Q}_{\kappa,h}|N_{h},\bar d\to q_\text{hard}) \, p(\bar d\to q_\text{hard}|c_W)+p({\cal Q}_{\kappa,h}|N_{h}, u\to q_\text{hard}) \, p( u\to q_\text{hard}|c_W)
\right] \nonumber\\
&
\hspace{-1cm}
\times\left[
p({\cal Q}_{\kappa,s}|N_{s},\bar d\to q_\text{soft}) \, p(\bar d\to q_\text{soft}|c_W)+p({\cal Q}_{\kappa,s}|N_{s}, u\to q_\text{soft}) \, p( u\to q_\text{soft}|c_W)
\right], 
\end{align}
where $p({\cal Q}|N,\bar d)$ and $p({\cal Q}|N,u)$ take the Gaussian forms from Eqs.~\eqref{eq:gauss1} and \eqref{eq:gauss2}.

\begin{figure}[!t]
    \centering
    \includegraphics[width=0.49\linewidth,clip]{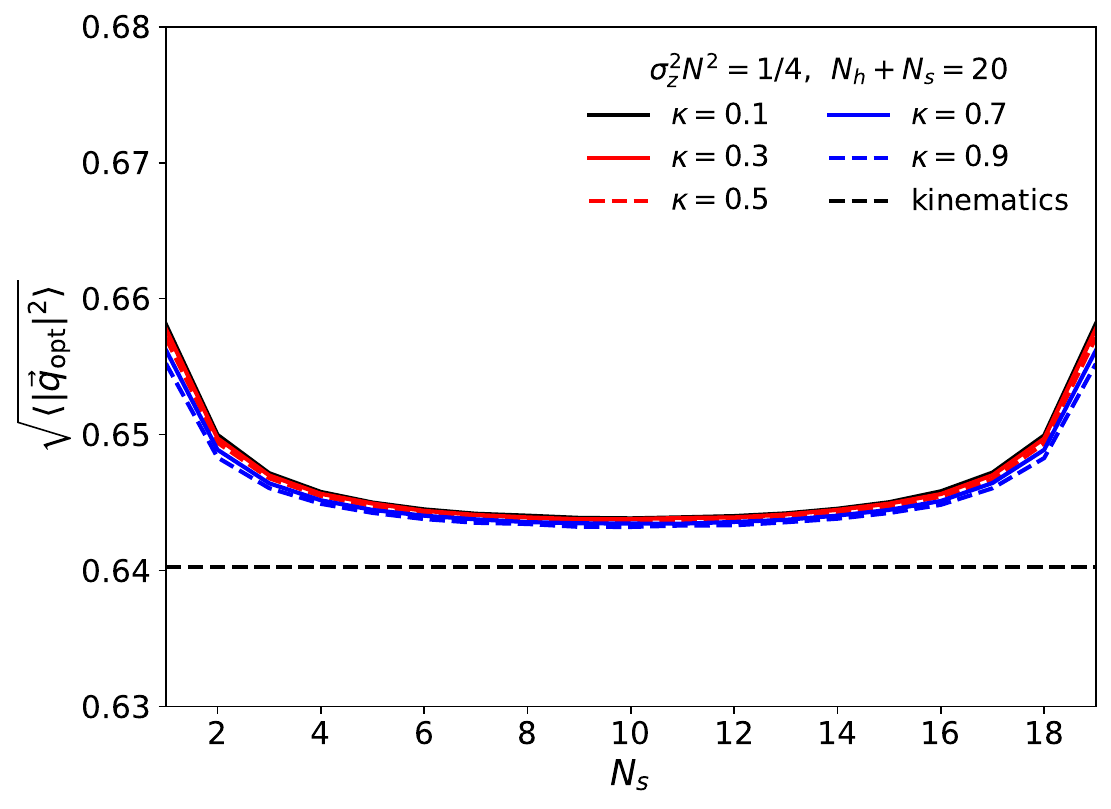} \hspace*{0.1cm}
    \includegraphics[width=0.49\linewidth,clip]{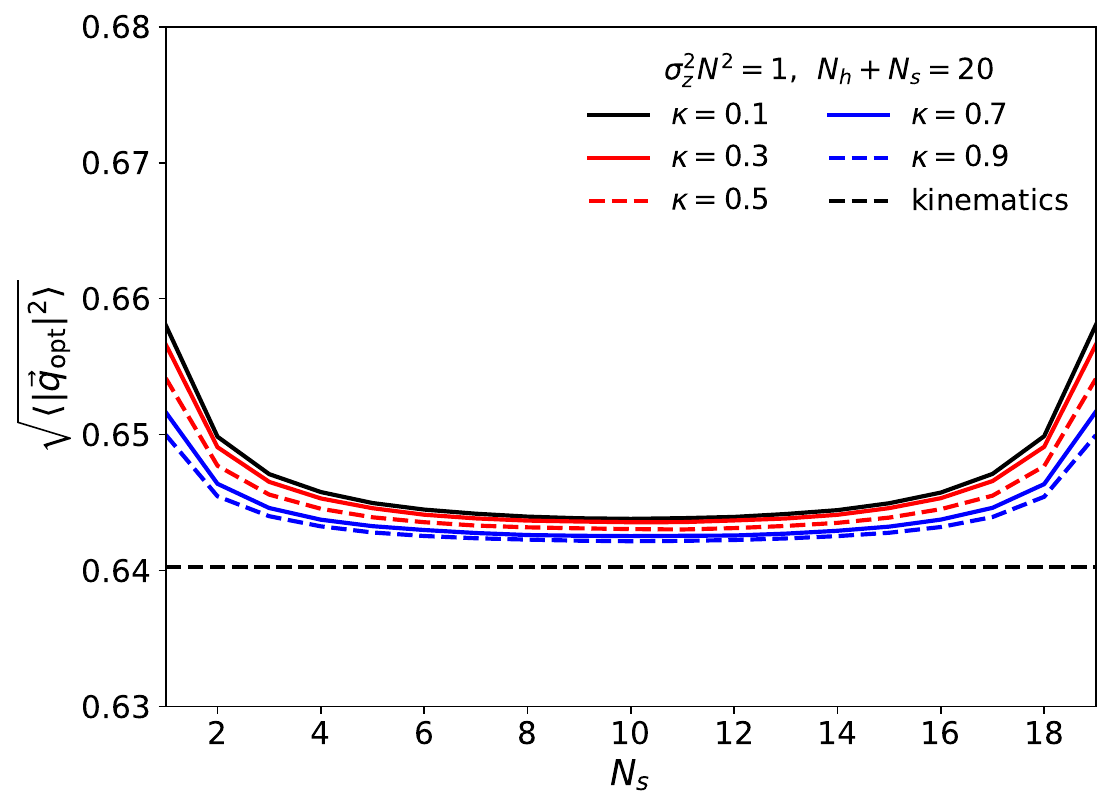} \\
    \caption{Spin analyzing power $\sqrt{\langle|\vec{q}_{\text{opt}}|^{2} \rangle}$ as a function of the particle multiplicity $N_s$, assuming $N_h+N_s=20$. We present the results for different values of $\kappa$, taking $\sigma_{z}^{2}N^{2}=1/4$ (left panel) and $\sigma_{z}^{2}N^{2}=1$ (right panel).}
    \label{fig:sqrt_qopt_sq}
\end{figure} 

\begin{figure*}[!t]
	\centering 
	\hspace{0.2cm}
    \includegraphics[width=0.48\textwidth]{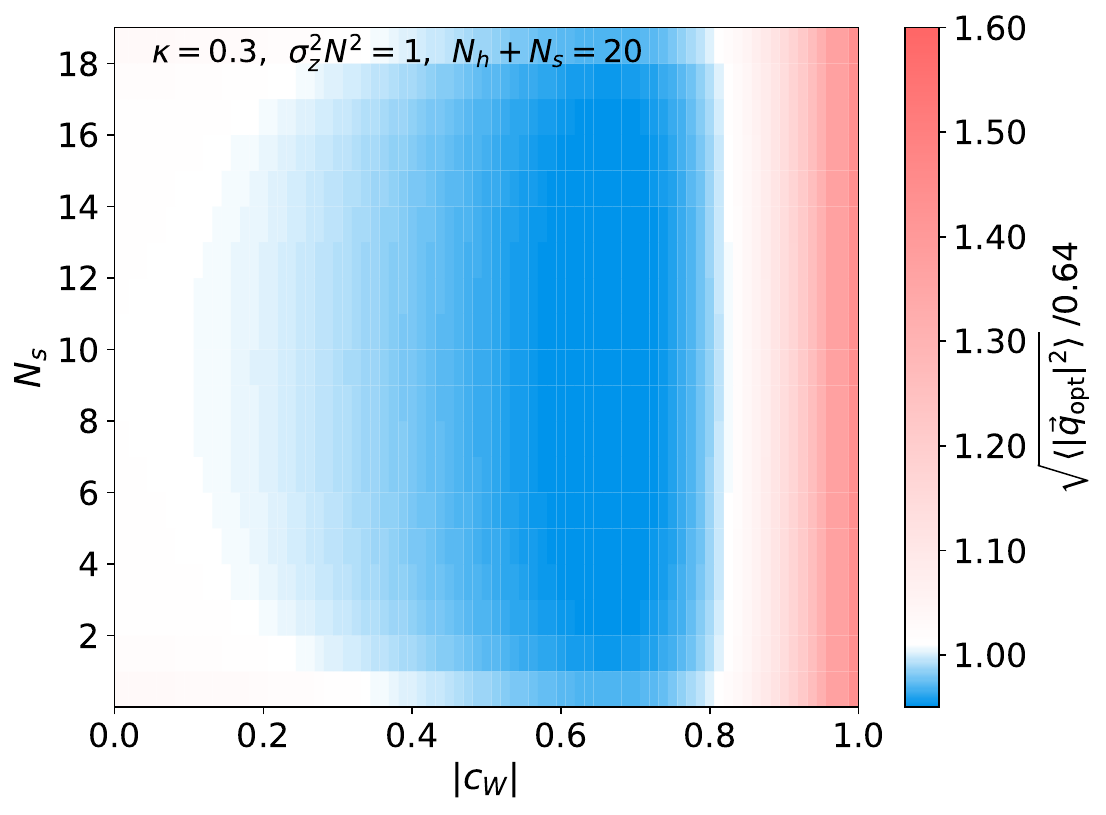} \hspace*{-0.2cm}
	\includegraphics[width=0.49\textwidth]{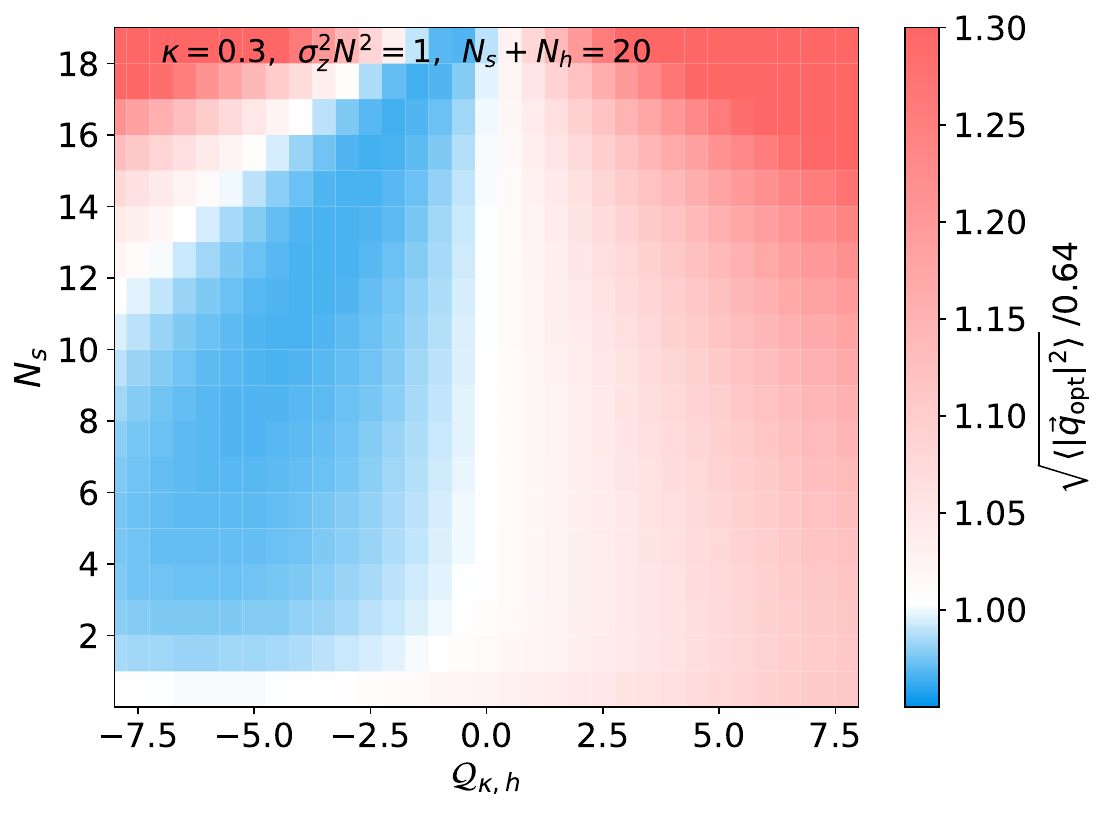} \\
     \includegraphics[width=0.49\textwidth]{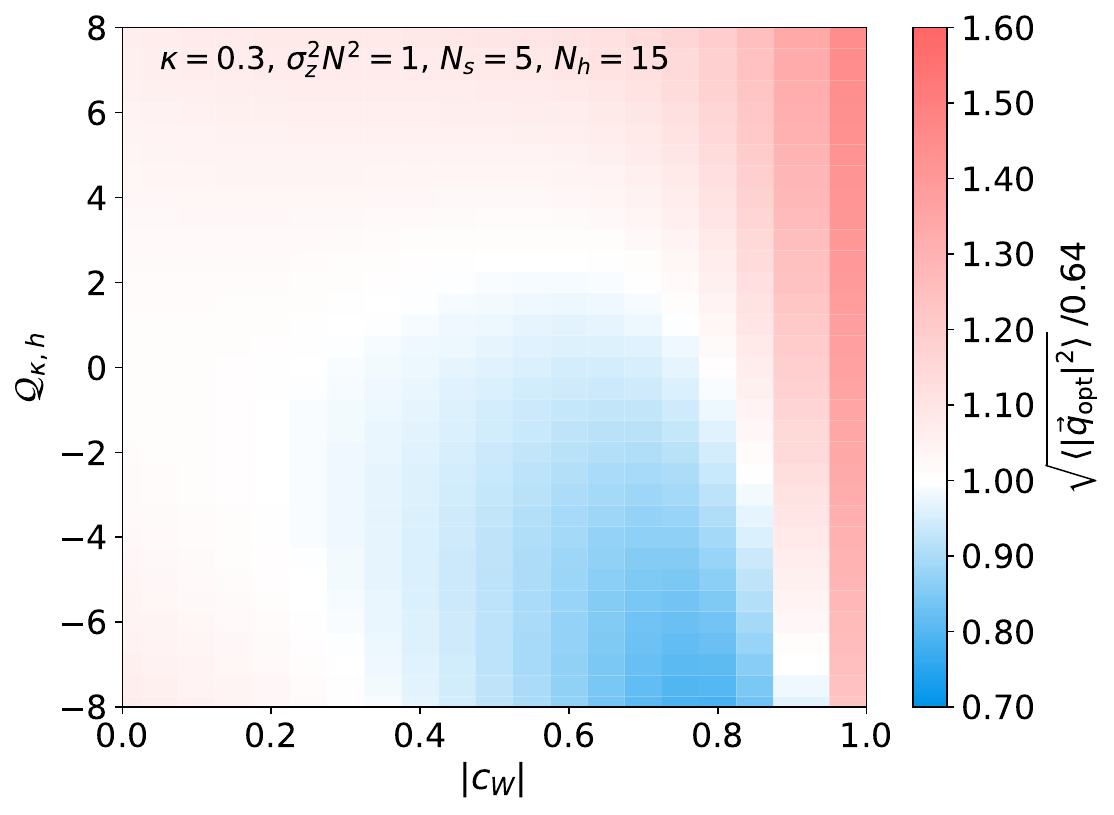}\hspace*{-0.cm}
     \includegraphics[width=0.49\textwidth]{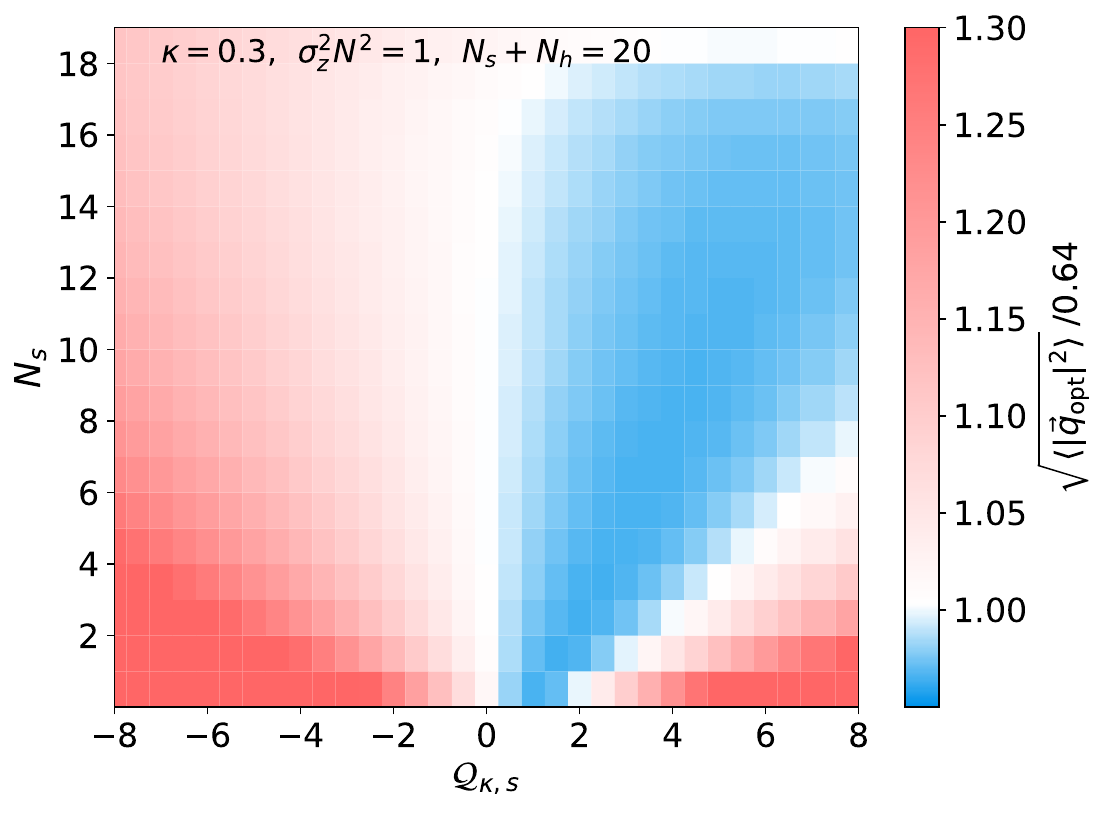} \\
	\includegraphics[width=0.49\textwidth]{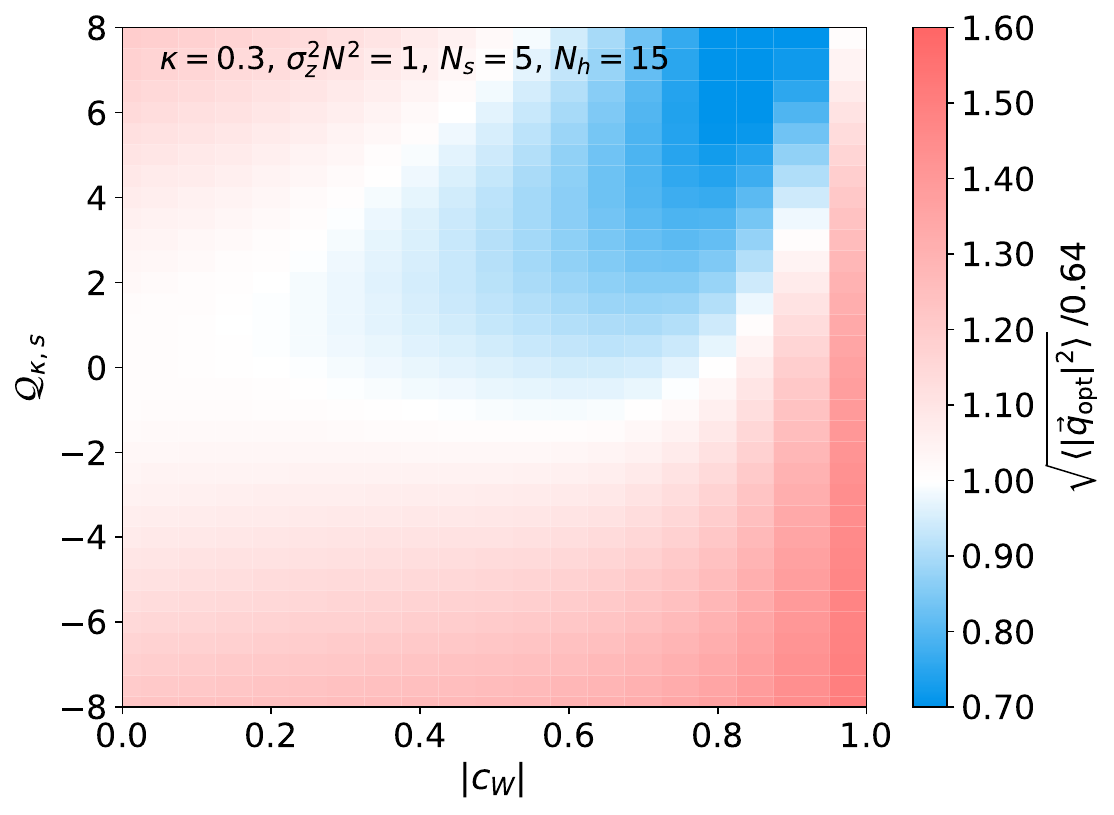}\hspace*{-0.cm}
    \includegraphics[width=0.49\textwidth]{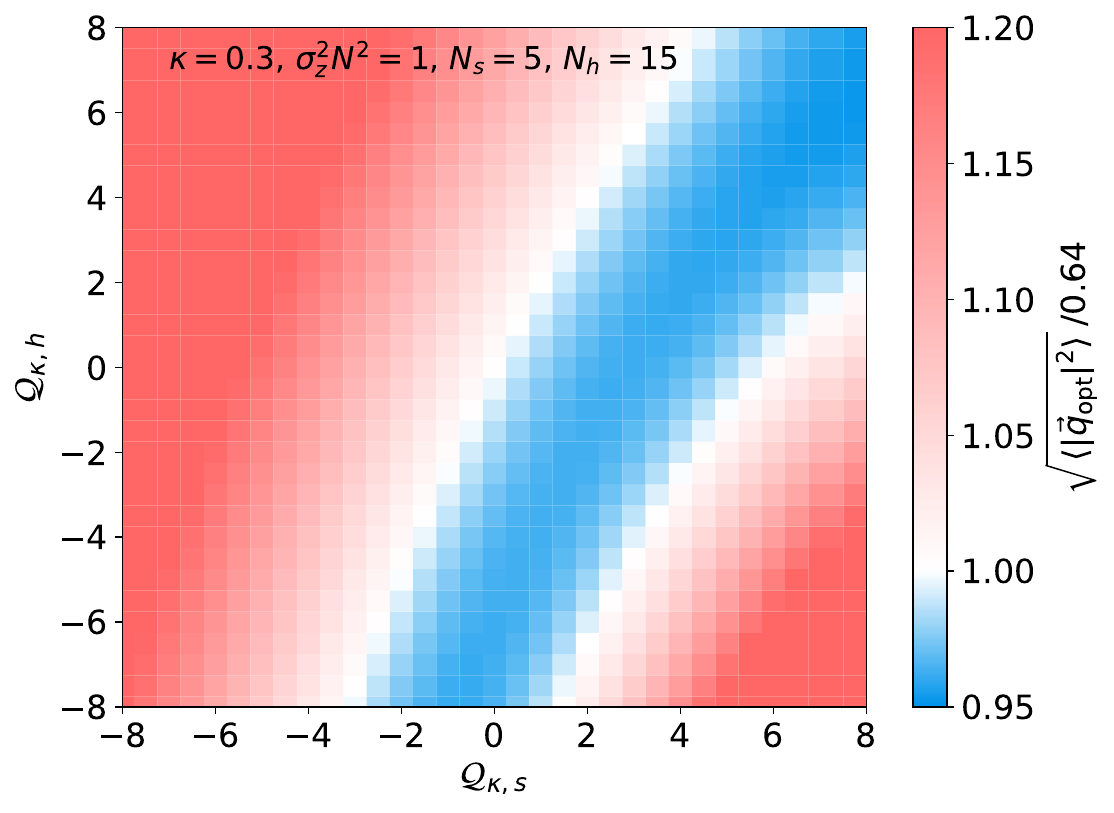}
    \caption{Square root of the mean square optimal direction as a function of multiplicity and helicity angle (upper left), jet charge of the hard jet and helicity angle (middle left), jet charge of the soft jet and helicity angle (bottom left), jet charge of the hard jet and multiplicity (upper right), jet charge of the soft jet and multiplicity (middle right), and jet charge of the hard and soft jets (bottom right). We assume $\kappa=0.3$ and $\sigma_{z}^{2}N^{2}=1$.} 
	\label{fig:sqrt_qopt_sq_2d}
\end{figure*}

The integrand in Eq.~\eqref{eq:qoptmeansq} is a complicated function of ${\cal Q}_{\kappa,h}$, ${\cal Q}_{\kappa,s}$ and $c_{W}$, so we perform the numerical integration for a fixed particle multiplicity $N_s$, assuming $N_s+N_h=20$, and different values of $\kappa$.
To illustrate the dependence on the width, we will present results with $\sigma_{z}^{2}N^{2}=(1/4,~1)$.
Our results are presented in Fig.~\ref{fig:sqrt_qopt_sq}, which shows that $\sqrt{\langle |\vec{q}_{\text{opt}}|^{2} \rangle}$ presents a U-shape profile due to the fixed $N_{s}+N_{h}$. The spin analyzing power increases for small $N_{s}$ with large $N_{h}$, as well as large $N_{s}$ with small $N_{h}$. In addition, we note that the spin analyzing power slightly increases for smaller $\kappa$.  This is in agreement with the results reported in Ref.~\cite{Kang:2023ptt}, indicating an increase in quark discrimination power as $\kappa$ approaches zero. 

To further investigate a possible enhancement of the spin analyzing power, we explore two-dimensional correlations between the observables: multiplicity of the soft jet $N_s$, the charges of the hard and soft jets $({\cal Q}_{\kappa,h},{\cal Q}_{\kappa,s})$, and helicity angle $c_W$. For this purpose, we defined the average of the squared-vector length over a region, ${\cal R}$, 
\begin{align}
    \langle |\vec{q}_{\text{opt}}|^{2} \rangle_{{\cal R}} =\frac{ \int_{{\cal R}} \, dc_{W}\, d{\cal Q}_{\kappa,h}\, d{\cal Q}_{\kappa,s}\, p(c_{W})\,p({\cal Q}_{\kappa,h},{\cal Q}_{\kappa,s}|c_{W},N_{h},N_{s})\,|\vec{q}_{\text{opt}}|^{2}}{\int_{{\cal R}} \, dc_{W}\, d{\cal Q}_{\kappa,h}\, d{\cal Q}_{\kappa,s}\, p(c_{W})\,p({\cal Q}_{\kappa,h},{\cal Q}_{\kappa,s}|c_{W},N_{h},N_{s})}, 
\end{align} 
where $\cal R$ denotes each bin of the two-dimensional distribution.
The corresponding results are presented in Fig.~\ref{fig:sqrt_qopt_sq_2d}. We observe that the spin analyzing power increases for large helicity angles, getting close to unity as $c_W$ approaches 1. In this limit, the down quark rarely decays collinear to the $W$ boson in the top quark rest frame because $f_R\simeq 0$ in Eq.~\eqref{eq:polar}. Hence, it is possible to obtain the full spin analyzing power in this kinematic regime. Furthermore, there is a significant increase in spin analyzing power for particular jet charge values. More concretely, we observe an enhancement for large and positive ${\cal Q}_{\kappa,h}$ (large and negative ${\cal Q}_{\kappa,s}$). This phenomenological behavior was anticipated in the discussion after Eq.~\eqref{eq:pdhard}. These features can be easily incorporated as a part of event selection at the cost of statistics, taking advantage of the large branching fraction for the hadronic top quark final state.

 \begin{figure*}[!t]
 	\centering 
\includegraphics[width=0.33\textwidth]{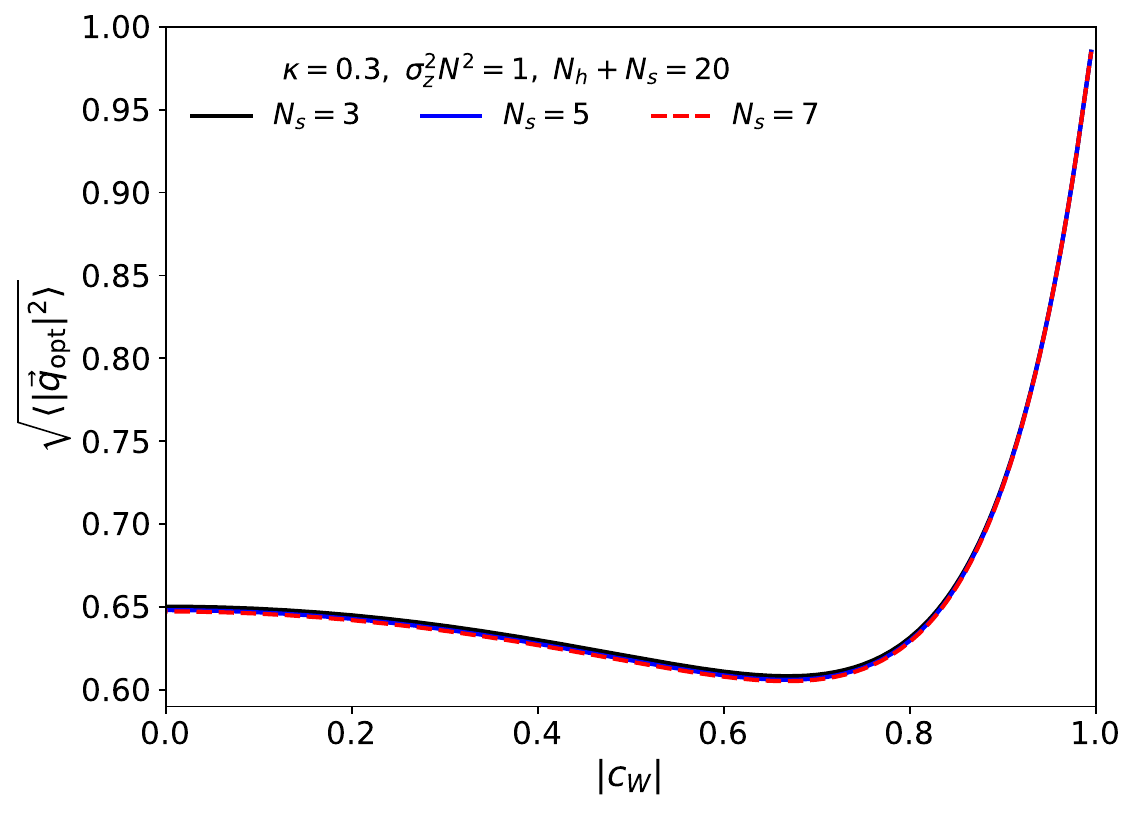} \hspace*{-0.2cm}
\includegraphics[width=0.33\textwidth]{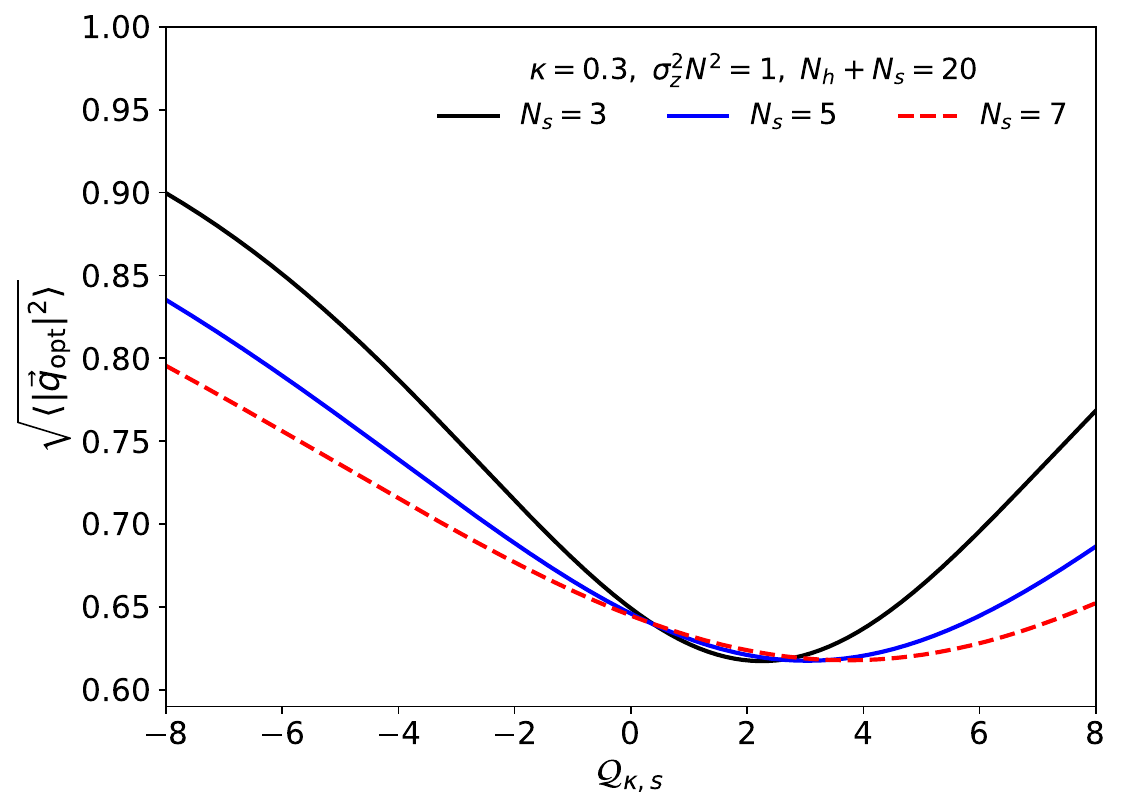} \hspace*{-0.2cm}
\includegraphics[width=0.33\textwidth]{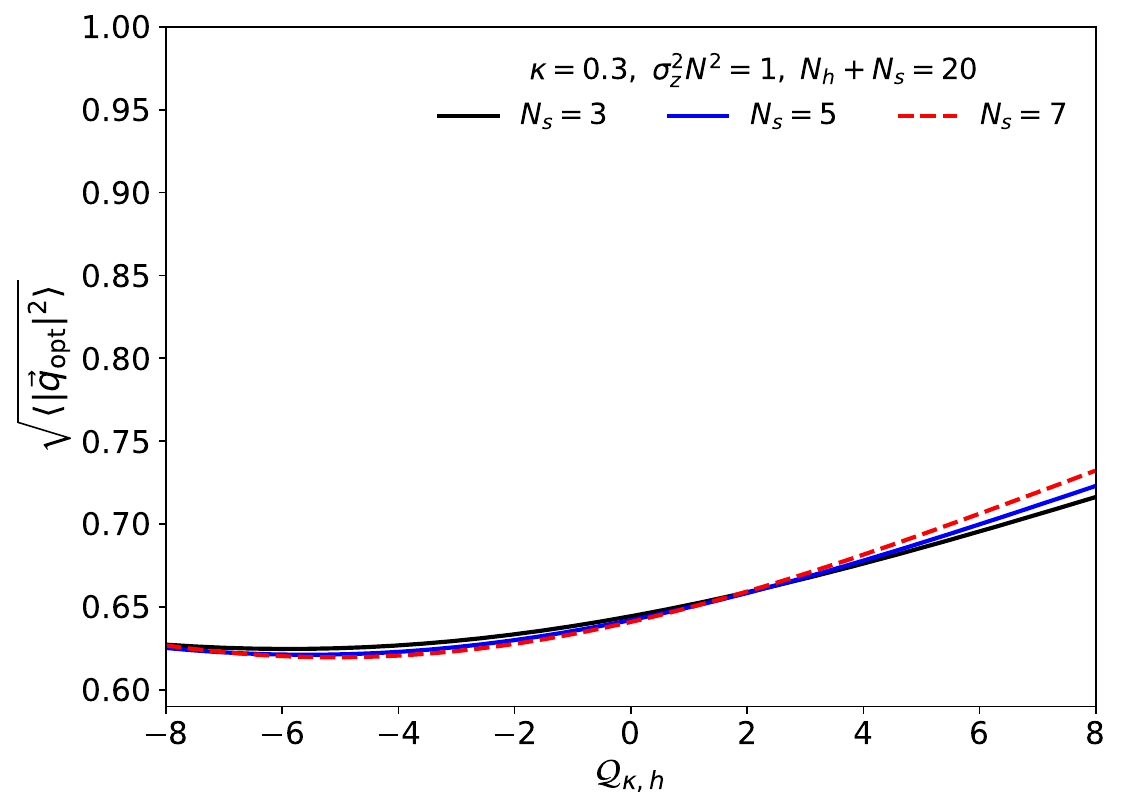} \\  
\includegraphics[width=0.33\textwidth]{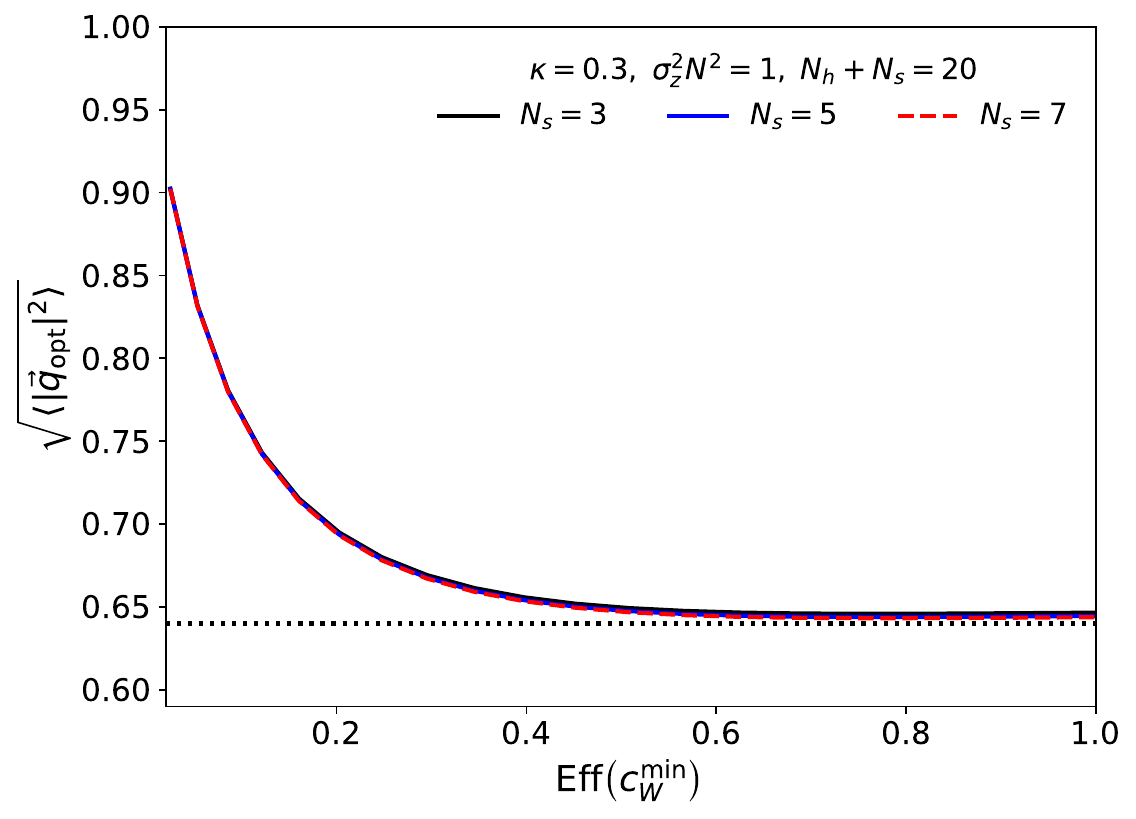} \hspace*{-0.2cm}
\includegraphics[width=0.33\textwidth]{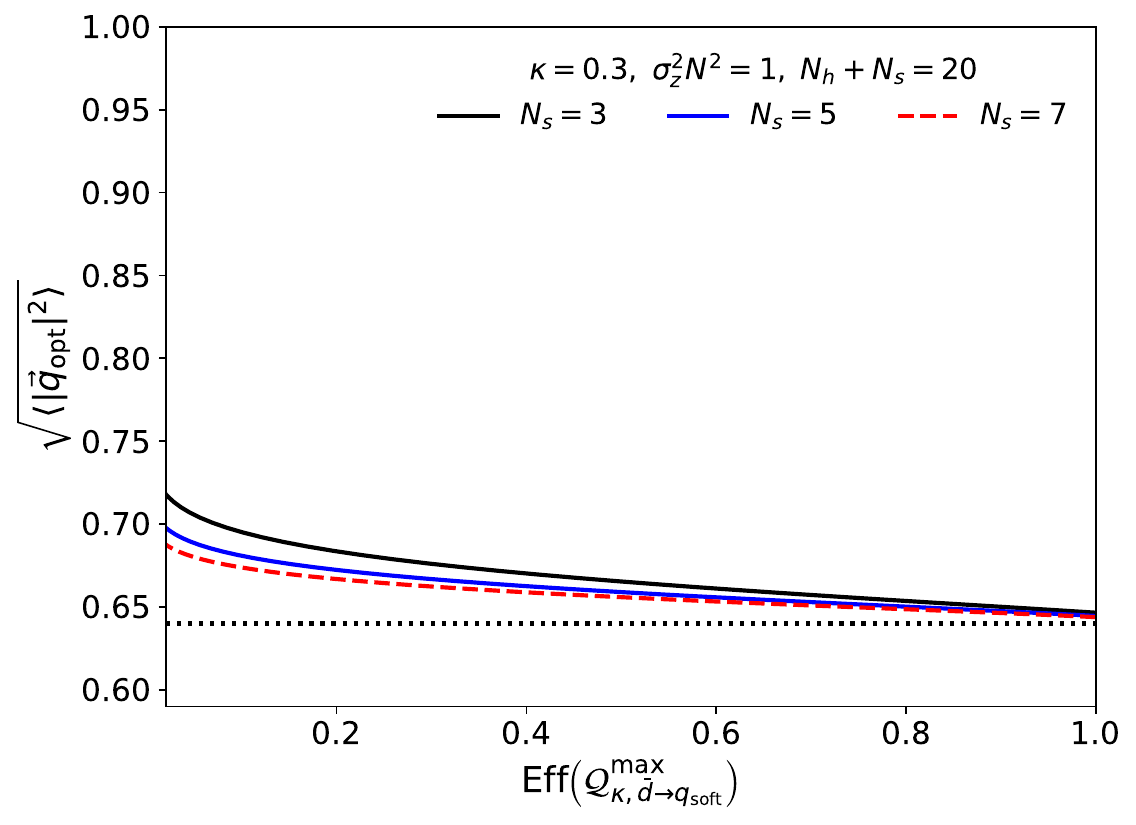} \hspace*{-0.2cm}
\includegraphics[width=0.33\textwidth]{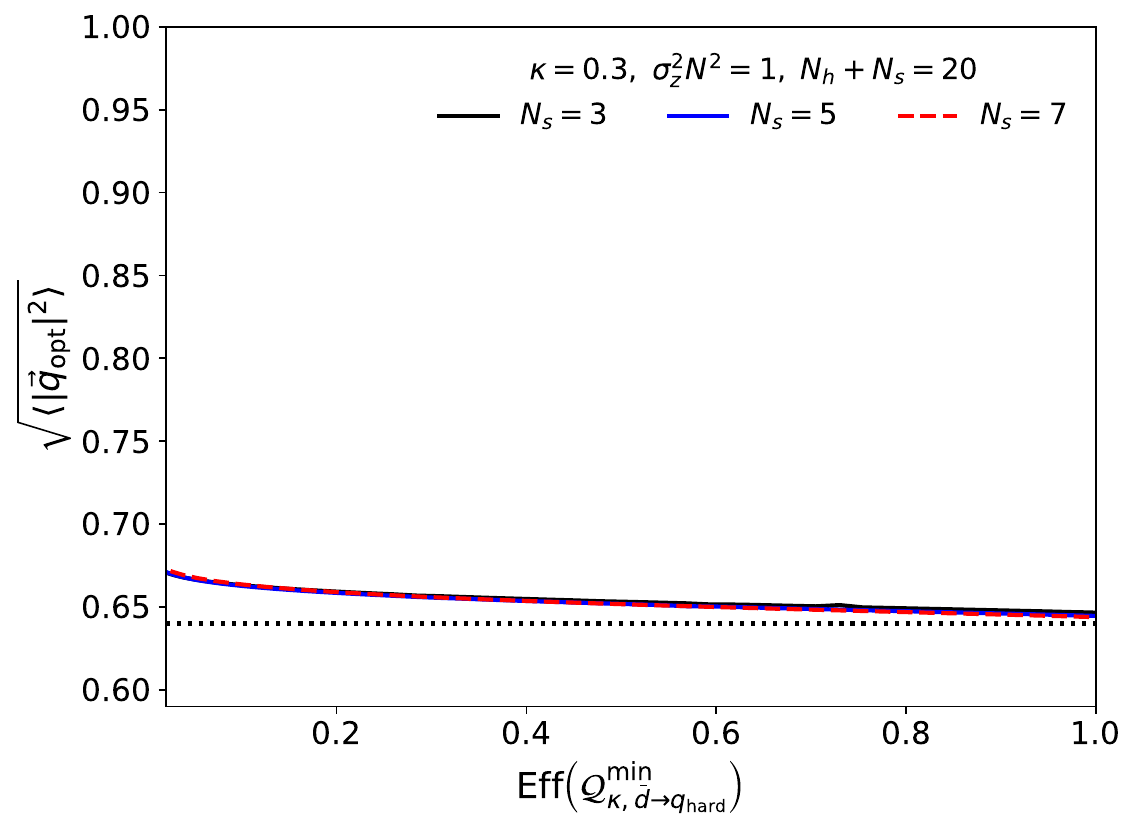} 
\caption{In the top panel, we present the average of the spin analyzing power $\sqrt{\langle|\vec{q}_{\text{opt}}|^{2} \rangle}$ as a function of the helicity angle $c_W$ (left), jet charge for the soft jet ${\cal Q}_{\kappa,s}$ (middle), and hard jet ${\cal Q}_{\kappa,h}$ (right). The results are shown for different multiplicities of the soft jet $N_s=3$ (black), $N_{s}=5$ (blue), and $N_{s}=7$ (red), assuming $N_s+N_h=20$. In the bottom panel, we show the average of the spin analyzing power for distinct efficiency cuts on $c_W$ (left),  ${\cal Q}_{\kappa,\bar d \to q_\text{soft}}^\text{max}$ (middle), and ${\cal Q}_{\kappa,\bar d\to q_\text{hard}}^\text{min}$ (right). We assume $\kappa=0.3$ and $\sigma_{z}^{2}N^{2}=1$.}
\label{fig:eff}
\end{figure*}

\begin{table}[!tb]
\renewcommand\arraystretch{1.3}
  \begin{tabular}{l||c|cc|cc}
    \hline
      & $\text{Eff}=100\%$
      & $\text{Eff}(c_W^{\rm min})=50\%$
      & $\text{Eff}(c_W^{\rm min})=20\%$
      & $\text{Eff}({\cal Q}_{\kappa,s}^{\rm max})=50\%$
      & $\text{Eff}({\cal Q}_{\kappa,s}^{\rm max})=20\%$
      \\
      \hline
$\sqrt{\langle|\vec{q}_{\text{opt}}|^{2} \rangle}_{N_s=3}$ & ~~0.646~~ & ~~0.649~~ & ~~0.695~~& ~~0.666~~ & ~~0.683~~ \\
$\sqrt{\langle|\vec{q}_{\text{opt}}|^{2} \rangle}_{N_s=5}$ & ~~0.644~~ & ~~0.647~~ & ~~0.694~~& ~~0.659~~ & ~~0.672~~ \\
$\sqrt{\langle|\vec{q}_{\text{opt}}|^{2} \rangle}_{N_s=7}$ & ~~0.644~~ & ~~0.647~~ & ~~0.693~~& ~~0.655~~ & ~~0.667~~ \\
\hline
  \end{tabular}
   \caption{Spin analyzing power $\sqrt{\langle|\vec{q}_{\text{opt}}|^{2} \rangle}$ for different efficiencies and soft jet multiplicities $N_s=(3,5,7)$.}
\label{table:betas}
\end{table}

In Fig.~\ref{fig:eff}, we illustrate the magnitude of this improvement. The top panel displays the spin analyzing power as a function of helicity angle, and jet charges for the soft and hard jets. The results are shown for distinct soft jet multiplicities $N_s=(3,5,7)$, assuming $N_s+N_h=20$. While the average of the spin analyzing power as a function of the helicity angle does not show significant changes for different $N_s$ values (and  the changes for ${\cal Q}_{\kappa,h}$ version of this plot are only mild), the average of the spin analyzing power as a function of the jet charge ${\cal Q}_{\kappa,s}$ shows relevant changes for different $N_s$. Based on the profiles of these distributions, we define minimal or maximal thresholds in the phase space to enhance the spin analyzing power. This is presented in terms of the efficiency of the corresponding selection, shown in the bottom panel of the same figure and illustrated in Table~\ref{table:betas} for particular efficiency values. We observe that by implementing a minimal threshold on the helicity angle, $c_W>c_W^{\rm min}$, it is possible to increase the spin analyzing power from 0.64 by 1.4\% (8.6\%) for 0.5 (0.2) efficiency. Similarly, we examine the potential enhancement arising from the jet charge distributions. In particular, for ${\cal Q}_{\kappa,\bar d\to \rm q_{soft}}<{\cal Q}_{\kappa,\bar d\to \rm q_{soft}}^{\rm max}$, the possible increase can reach up to approximately 4.0\% (6.7\%) for 0.5 (0.2) efficiency. Hence, considering the studied observables, the major gains for the hadronic top quark polarimetry will come from a combination of $c_W$ and ${\cal Q}_{\kappa,s}$ dependencies.

\section{Conclusion}
\label{sec:conclusion}
 
The measurement of top quark polarization provides a unique and powerful tool to advance precision physics and explore new physics beyond the Standard Model. While its leptonic decay serves as a clean proxy for the top quark polarimetry, it comes with significant statistical limitations. In contrast,  the sizable hadronic decays of the top quark, despite their complexity, offer a promising avenue for polarization studies~\cite{Tweedie:2014yda,PhysRevD.109.115023,Dong:2024xsg,CMS:2024vqh}.

In this paper, we investigated the relevance of global jet dynamics -- arising from kinematics, jet charges, and particle multiplicity -- for enhancing the sensitivity to hadronic top quark polarimetry. By applying the general formalism presented in Ref.~\cite{Dong:2024xsg}, we provided analytic parametric estimates of the improvement in the spin analyzing power for hadronic top quarks. Our analysis demonstrated that the considered observables can lead to substantial improvements in specific regions of phase space. In particular, we observed that the helicity angle and jet charge for the soft jet yield sizable effects, underscoring their importance in hadronic top quark polarimetry. These analytical derivations provide deeper insights into the sources of improvement in hadronic top quark polarimetry, building on the foundation laid in Ref.~\cite{Dong:2024xsg}.

\medskip
\medskip
\noindent {\bf Acknowledgments.}
The authors thank the organizers for the Mitchell Conference on Collider, Dark Matter, and Neutrino Physics held at Texas A\&M University, where the inspiration for this project originated.
D.G. thanks the group at the IPPP-Durham University for hosting him during the final stages of this project.
DG and AN thank the U.S.~Department of Energy for the financial support, under grant number DE-SC 0016013. 
A.L. was supported in part by the UC Southern California Hub, with funding from the UC National Laboratories division of the University of California Office of the President.
K.K. is supported by US DOE DE-SC0024407 and  Z.D. is supported in part by US DOE DE-SC0024673 and by College of Liberal Arts and Sciences Research Fund at the University of Kansas.
K.K. would like to thank the Aspen Center for Physics and the organizers of Summer 2024 workshop, ``Fundamental Physics in the Era of Big Data and Machine Learning'' (supported by National Science Foundation grant PHY-2210452) for hospitality during the completion of this manuscript.

\appendix

\section{Enhancing Spin Analyzing Power with Jet Charge: a Simple Proof}
\label{app:impro}

In this appendix, we provide a simple proof that the spin analyzing power is always improved by
incorporating jet charge information. The spin analyzing power can be examined in terms of the length of the vector
\begin{align}
\vec q_\text{opt} &= p(\bar d\to q_\text{hard}|c_W,{\cal Q}_{\kappa,h},N_h,{\cal Q}_{\kappa,s},N_s) \, \hat q_\text{hard} +p(\bar d\to q_\text{soft}|c_W,{\cal Q}_{\kappa,h},N_h,{\cal Q}_{\kappa,s},N_s) \, \hat q_\text{soft}\,.
\end{align}
Then, $|\vec q_\text{opt}|^2$ becomes 
\begin{align}
|\vec q_\text{opt}|^2 &= 1-2x(1-x)(1-\hat q_\text{hard}\cdot \hat q_\text{soft})\,,
\end{align}
where $x\equiv p(\bar d\to q_\text{hard}|c_W,{\cal Q}_{\kappa,h},N_h,{\cal Q}_{\kappa,s},N_s)$.
For a fixed value of the dot product $\hat q_\text{hard}\cdot \hat q_\text{soft}$, this length increases if the product $x(1-x)$ decreases.  With this in mind, we can analyze how the inclusion of jet charge improves the spin resolving power.

Assuming that the measured value of $c_W$ is such that $p(\bar d\to q_\text{hard}|c_W) < 1/2$, we will argue that $p(\bar d\to q_\text{hard}|c_W,{\cal Q}_{\kappa,h},N_h,{\cal Q}_{\kappa,s},N_s)<p(\bar d\to q_\text{hard}|c_W)$, and correspondingly that the spin analyzing power has improved by measuring the jet charge information.  Consider the first factor of $p(\bar d\to q_\text{hard}|c_W,{\cal Q}_{\kappa,h},N_h,{\cal Q}_{\kappa,s},N_s)$ in Eq. \eqref{eq:pdhard}
\begin{align}
\frac{1}{p(\bar d\to q_\text{hard}|c_W)+\frac{p({\cal Q}_{\kappa,h}| u\to q_\text{hard},N_h)}{p({\cal Q}_{\kappa,h}|\bar d\to q_\text{hard},N_h)}\,p( u\to q_\text{hard}|c_W)}\,.
\end{align}
Necessarily, we have the sum rule, $p(\bar d\to q_\text{hard}|c_W)+p(u\to q_\text{hard}|c_W)=1$.  Because we assume that $p(\bar d\to q_\text{hard}|c_W) < 1/2$, most of the time the harder jet is $ u$, and so on average the ratio
\begin{align}
\frac{p({\cal Q}_{\kappa,h}| u\to q_\text{hard},N_h)}{p({\cal Q}_{\kappa,h}|\bar d\to q_\text{hard},N_h)} > 1\,.
\end{align}
Therefore, the first factor is less than 1: 
\begin{align}
\frac{1}{p(\bar d\to q_\text{hard}|c_W)+\frac{p({\cal Q}_{\kappa,h}| u\to q_\text{hard},N_h)}{p({\cal Q}_{\kappa,h}|\bar d\to q_\text{hard},N_h)}\,p( u\to q_\text{hard}|c_W)}<1\,.
\end{align}
A similar argument holds for the second factor in Eq.~\eqref{eq:pdhard}, and we have 
\begin{align}
\frac{1}{\frac{p({\cal Q}_{\kappa,s}|\bar d\to q_\text{soft},N_s)}{p({\cal Q}_{\kappa,s}| u\to q_\text{soft},N_s)}\,p(\bar d\to q_\text{soft}|c_W)+p( u\to q_\text{soft}|c_W)} < 1\,.
\end{align}
Thus, we necessarily find that $p(\bar d\to q_\text{hard}|c_W,{\cal Q}_{\kappa,h},N_h,{\cal Q}_{\kappa,s},N_s)<p(\bar d\to q_\text{hard}|c_W)$.  Similarly, it is also true that $p(\bar d\to q_\text{soft}|c_W,{\cal Q}_{\kappa,h},N_h,{\cal Q}_{\kappa,s},N_s)>p(\bar d\to q_\text{soft}|c_W)$ for the same value of $c_W$.  

Now, $x(1-x)$ is maximized where $x=1/2$, but we assumed that we started away from the maximum and with the inclusion of jet charge moved further from the maximum because the difference $p(\bar d\to q_\text{soft}|c_W,{\cal Q}_{\kappa,h},N_h,{\cal Q}_{\kappa,s},N_s) - p(\bar d\to q_\text{hard}|c_W,{\cal Q}_{\kappa,h},N_h,{\cal Q}_{\kappa,s},N_s)$ is larger than $p(\bar d\to q_\text{soft}|c_W)-p(\bar d\to q_\text{hard}|c_W)$.  Therefore, the spin analyzing power must be improved by measuring jet charge information.

\bibliographystyle{unsrt}
\bibliography{draft}

\end{document}